\pdfoutput=1

\documentclass[11pt]{article}

\usepackage[]{acl}

\usepackage{times}
\usepackage{latexsym}

\usepackage{graphicx}
\usepackage{xurl}

\usepackage{amsmath, xparse}
\usepackage{booktabs}
\usepackage{array}
\usepackage{makecell}
\usepackage{subfig}
\usepackage{tabularx}
\usepackage{multirow}

\usepackage{algorithm}
\usepackage{algorithmic}
\usepackage{newfloat}
\usepackage{listings}
\usepackage{stfloats}
\usepackage{ulem}

\def\arraystretch{1.4}
\usepackage[T1]{fontenc}

\usepackage[utf8]{inputenc}

\usepackage{microtype}

\title{Leveraging Unimodal Self-Supervised Learning for\\Multimodal Audio-Visual Speech Recognition}

\author{Xichen Pan$^1$, Peiyu Chen$^1$, Yichen Gong$^2$, Helong Zhou$^2$, Xinbing Wang$^1$, Zhouhan Lin$^1$\footnotemark[2] \\
   $^1$Shanghai Jiao Tong University, $^2$Horizon Robotics\\
   \small \texttt{\{flash321, pietychen, xwang8\}@sjtu.edu.cn, \{yichen01.gong, helong.zhou\}@horizon.ai}\\
   \small \texttt{lin.zhouhan@gmail.com}
}

\begin{document}
\maketitle

\renewcommand{\thefootnote}{\fnsymbol{footnote}} 
\footnotetext[2]{Corresponding author.}

\begin{abstract}
Training Transformer-based models demands a large amount of data, while obtaining aligned and labelled data in multimodality is rather cost-demanding, especially for audio-visual speech recognition (AVSR). Thus it makes a lot of sense to make use of unlabelled unimodal data. On the other side, although the effectiveness of large-scale self-supervised learning is well established in both audio and visual modalities, how to integrate those pre-trained models into a multimodal scenario remains underexplored. In this work, we successfully leverage unimodal self-supervised learning to promote the multimodal AVSR. In particular, audio and visual front-ends are trained on large-scale unimodal datasets, then we integrate components of both front-ends into a larger multimodal framework which learns to recognize parallel audio-visual data into characters through a combination of CTC and seq2seq decoding. We show that both components inherited from unimodal self-supervised learning cooperate well, resulting in that the multimodal framework yields competitive results through fine-tuning. Our model is experimentally validated on both word-level and sentence-level tasks. Especially, even without an external language model, our proposed model raises the state-of-the-art performances on the widely accepted Lip Reading Sentences 2 (LRS2) dataset by a large margin, with a relative improvement of 30\%. \footnote{Our codes are available at \url{https://github.com/LUMIA-Group/Leveraging-Self-Supervised-Learning-for-AVSR}.} 


\end{abstract}

\section{Introduction}
Audio-Visual Speech Recognition (AVSR) is a speech recognition task that leverages both an audio input of human voice and an aligned visual input of lip motions. It has been one of the successful application fields that involve multiple modalities in recent years. Due to the limited amount of labeled, multimodal aligned data and the difficulty of recognition from the visual inputs (i.e., lip reading), it is a challenging task to tackle. 

Existing AVSR models tend to use extra data to increase the performance of the system, in a form of inserting an extra supervised learning stage in the training process. For example, many existing methods rely on an extra sequence level classification to bootstrap its learning on visual features. \citet{avsrhybrid, convseq2seq} train their visual front-end on LRW \cite{LRW} before learning on the AVSR task. \citet{deepavsr, LRS3} chunks the MV-LRS data \cite{chung2017lip} into pieces of words and pre-train the model through classification. VoxCeleb \cite{chung2018voxceleb2} are also used in \citet{afouras2020asr} for the same purpose. Learning an effective visual front-end could still be notoriously hard, even with these extra supervised learning tasks. Sometimes curriculum learning is required to adapt the learned visual front-end into AVSR task \cite{deepavsr}. End-to-end learning of large-scale AVSR data hasn't been successful until recently \cite{e2econformer}. 

    Although self-supervised learning could enable leveraging unlabelled or even unaligned data, it hasn't been adequately explored on this task. \citet{shukla2020visually} is among the few attempts in this facet, in which it predicts lip motions from audio inputs. Their proposed learning schemes yield strong emotion recognition results but are relatively weak in speech recognition. Moreover, since in AVSR it is the lip shape and motions between frames rather than the objects in a single image that matters for recognizing speech contents, if pre-trained visual models tailored for tasks targeting at single frame images could work for AVSR remains unknown. In another scenario, self-supervised learning in unimodality has been well established as a paradigm to learn general representations from unlabelled examples, such as in natural language processing \cite{brown2020language, devlin2018bert}, speech recognition \cite{wav2vec2}, and computer vision \cite{mocov1, simclr, byol}.

In this work, we rely on a simple but effective approach, which is to utilize unlabelled unimodal data by using pre-trained models that are trained in single-modality through self-supervised learning. Specifically, we use \citet{wav2vec2} pre-trained on the large LibriLight \cite{librilight} dataset as our audio front-end. For visual front-end, we found that it is not as straight-forward for it to leverage pre-trained models, as we have to substitute the first convolutional layer in MoCo v2 \cite{mocov2} by a 3-D convolutional layer and fine-tune it through LRW. In total, our approach doesn't require a curriculum learning stage, and 
the overall training time has been decreased. 

Experimental results show that our new front-ends significantly outperform previous ones by a big margin in both audio-only and visual-only settings, and a new state-of-the-art has been achieved in the final AVSR setting. To our best knowledge, this is the first work that successfully applies unimodal pre-trained models in the multimodal setting of AVSR.

\section{Related Work}

\subsection{Audio-Visual Speech Recognition}
The earliest work on AVSR could be dated back to around two decades ago, when \citet{early_hmm_avsr} showed hand-crafted visual feature improves HMM-based ASR systems. The first modern AVSR system is proposed in \citet{deepavsr} where deep neural networks are used. The field has been rapidly developing since then. Most of the works are devoted into the architectural improvements, for example, \citet{convseq2seq} proposed temporal focal block and spatio-temporal fusion, and \citet{TMDCM} explored to use cross-modality attentions with Transformer. 

The other line of research focuses on a more diversified learning scheme to improve AVSR performance. \citet{li2019improving} uses a cross-modal student-teacher training scheme. \citet{paraskevopoulos2020multiresolution} proposes a multi-task learning scheme by making the model to predict on both character and subword level. Self-supervised learning has also been explored in \citet{shukla2020visually}, where the cross-modality setting is utilized by predicting frames of videos from audio inputs. 

The end-to-end learning of AVSR systems are first seen in \citet{tao2020end}, albeit in a much simpler dataset than LRS2. More recent work \cite{e2econformer} has made end-to-end learning on LRS2 possible by using a Conformer acoustic model and a hybrid CTC/attention decoder.

\subsection{Self-Supervised Learning}

Self-supervised learning has been chased in recent years since its ability to learn general representations of data through simple tasks that don't require labeling. Contrastive learning \cite{cl} has become the most impactful learning scheme in this field. In natural language processing, uni-or bi-directional language modelling \cite{brown2020language, devlin2018bert} have been used to significantly increase performances on various tasks. In audio speech processing, contrastive predictive coding \cite{wav2vec2} has been proven to be powerful in speech recognition. In the visual domain, Earlier works create self-supervised tasks through image processing based methods, such as distortion \cite{distortion},colorization \cite{colorization} and context prediction \cite{patches}. More recently, contrastive learning emerged as a paradigm of self-supervised learning, which results in a group of more expressive general visual representations, such as MoCo \cite{mocov1, mocov2}, SimCLR \cite{simclr}, BYOL \cite{byol}, etc.


\section{Architecture}
\begin{figure*}[t]
\centering
\includegraphics[width=\textwidth]{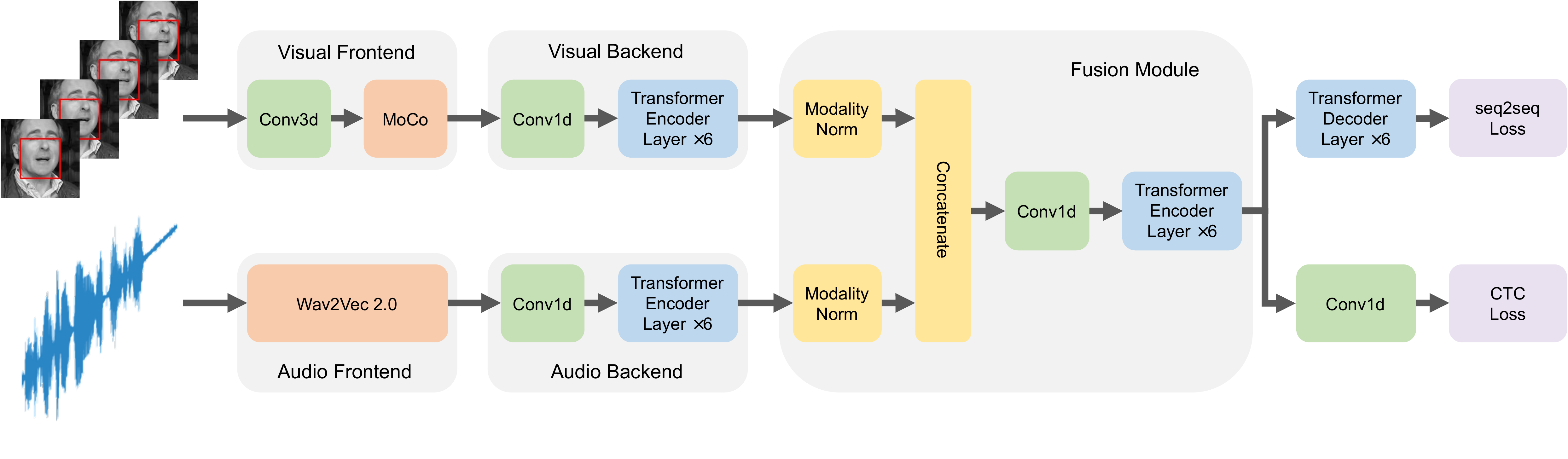}
\caption{Overall architecture of our AVSR model.}
\label{fig:framework}
\end{figure*}

The overall architecture of our model is shown in Fig. \ref{fig:framework}. The audio-visual model is comprised of four components, the front-ends and back-ends for both modalities, the fusion module, and the decoders. 

\subsection{Front-ends}
\noindent\textbf{Visual Front-end:}
Visual front-end serves as a component to capture the lip motion and reflect the lip position differences in its output representations. A naive way to apply pre-trained models in the visual front-end is to directly feed the RGB channels of each frame as input. However, since frames within a same clip in AVSR are largely similar in their contents while most pre-trained models in vision target at learning general representations reflecting the content of the whole image, this approach will result in similar outputs for all the frames, collapsing the informative lip position differences between frames.

\begin{figure*}[t]
\centering
\includegraphics[width=\textwidth]{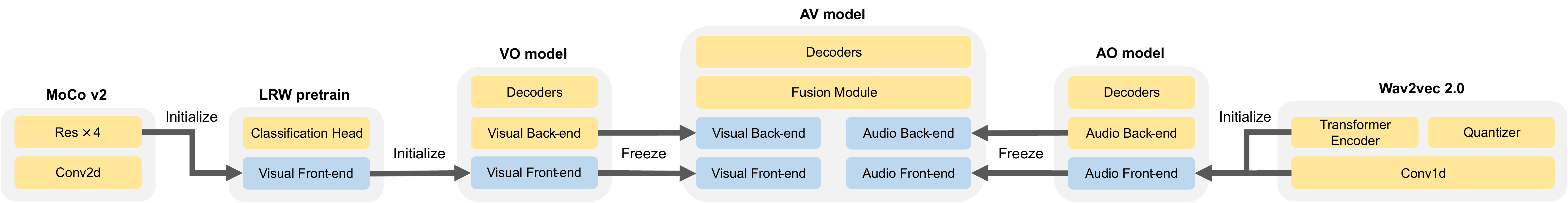}
\caption{Training pipeline of the model. Yellow blocks represent new parameters that are randomly initialized, while Blue blocks represent parameters that are inherited from last training stage.}
\label{fig:pipeline}
\end{figure*}

To overcome the aforementioned problem while still being able to utilize the pre-trained model, we truncate the first convolutional layer in MoCo v2 \cite{mocov2}, which is pre-trained on ImageNet \cite{imagenet}, and replace it with a layer of 3-D convolution. The outputs of 3-D convolutional layer are intentionally made identical to the input of the first ResBlock in MoCo v2 (see Table \ref{tb:visualdim}), thus providing a compatible interface to transfer higher layers of MoCo v2 into this task. On the other hand, we also adopt the common practice to convert the RGB input image to gray-scale before feeding it into the model, as it prevents the model from learning chromatic aberration information.\\[0.5em]
\noindent\textbf{Audio Front-end:}
The audio front-end is rather straight-forward. We use wav2vec 2.0 \cite{wav2vec} pre-trained on Libri-Light \cite{librilight}, like it is normally used for ASR tasks, both the 1-D convolutional layers and the stacked Transformer encoder layers are transferred into our audio front-end. The audio front-end takes as input raw audio wave of 16kHz, and produces one vector representation every 20ms. 
The audio feature dimensions are shown in Table \ref{tb:audiodim}.

\subsection{Back-ends}
Since the visual frames are in 25 FPS and the wav2vec 2.0 outputs are around 49 Hz, one should note that there is 2x difference in the frequency of frame-wise visual and audio representations at the output of their front-ends.\footnote{The odds are due to the larger receptive fields of wav2vec 2.0 1-D convolutional layers, which we circumvent by properly prefixing and suffixing the audio sequence and truncate the trailing audio vector. Thus a perfect 1:2 ratio of visual frames and audio front-end outputs are ensured. } In the back-end, we use 1-D convolutional layers on the time dimension combined with Transformer encoder layers to provide single modality temporal modeling, as well as adjusting the features to have the same frequency.
\\[0.5em]
\noindent\textbf{Visual Back-end:}
The incoming MoCo v2 output to the visual back-end has a feature dimension of 2048, at a frequency of 25 vectors per second. In the visual backend, we keep this frequency while reducing the feature size to 512. See Table \ref{tb:visualdim}. For positional encodings of the Transformer, we use fixed positional encoding in the form of sinusoidal functions. 

\noindent\textbf{Audio Back-end:}
In the audio back-end, the incoming wav2vec 2.0 outputs have a feature size of 1024, at a frequency of 50 vectors per second. We downscale the frequency by setting the stride of 1-D convolutional layer to 2. The Transformer encoder layers have the identical size to that of the visual back-end, while using a separate set of parameters. Table \ref{tb:audiodim} shows a clearer picture of audio front- and back-end dimensions.

\begin{table}[t]
    \small
    \centering
\begin{tabularx}{\linewidth}{c|c|X}
    \toprule
    Stage                    & Modules                & \makecell{Image sequence \\ $(T_f \times 112^2 \times 1)$}\\
    \hline
    \multirow{2}*{Front-end} & 3-D convolution  & \makecell{$(T_f \times 28^2 \times 64)$} \\
    \cline{2-3}
    ~                        & MoCo v2                & \makecell{$(T_f \times 2048)$}           \\
    \hline
    \multirow{2}*{Back-end}  & 1-D convolution & \makecell{$(T_f \times 512)$}            \\
    \cline{2-3}
    ~                        & Transformer encoder    & \makecell{$(T_f \times 512)$}            \\
    \bottomrule
\end{tabularx}
    \caption{The feature dimension of visual stream. The dimensions of features are denoted by $\{\text{temporal size} \times (\text{spatial size}^2) \times \text{channels}\}$. $T_f$ denotes the number of visual frames.}
    \label{tb:visualdim}
\end{table}

\begin{table}[t]
    \small
    \centering
\begin{tabularx}{\linewidth}{c|c|X}
    \toprule
    Stage                   & Modules                & \makecell{Audio waveform          \\ $(T_s \times 1)$}\\
    \hline
    Front-end               & wav2vec 2.0            & \makecell{$(T_f \times 1024)$}          \\
    \hline
    \multirow{2}*{Back-end} & 1-D convolution & \makecell{$(\frac{T_f}{2} \times 512)$} \\[0.1em]
    \cline{2-3}
    ~                       & Transformer encoder    & \makecell{$(\frac{T_f}{2} \times 512)$} \\
    \bottomrule
\end{tabularx}
    \caption{The feature dimension of audio stream. The dimensions of features are denoted by $\{\text{temporal size} \times \text{channels}\}$. $T_s$ and $T_f$ denote the number of sampled audio input and audio frames, respectively.}
    \label{tb:audiodim}
\end{table}


\subsection{Fusion Module}
Features from both the audio and visual modalities are fused together in this section, forming vector representation of 1024 dimensions at a relatively low rate of 25 Hz. We use LayerNorm \cite{layernorm} separately on each of the modalities before concatenating them on the feature dimension. The LayerNorm is required since it avoids one modality overtaking the whole representation with larger variance. Similar 1-D convolutional layers and a subsequent Transformer encoder block of 6 layers take the fused representations as input, and encode them for the decoders.


\subsection{Decoders}
Following the setting of \citet{avsrhybrid}, there are two decoders trained simultaneously based on the same output in the fusion module. 

The first is a Transformer seq2seq decoder, a Transformer decoder with 6 layers is used, and we perform teacher forcing at character level by using ground truth characters as input during training.

The second one is arguably a decoder since it yields character probabilities for each timestep and relies on the CTC loss in training. $4$ extra 1-D convolutional layers with ReLU activation are used on top of the last Transformer encoder layer output. We also include LayerNorm between each of the layers.




\subsection{Loss Functions}
In this work, we use a so called hybrid CTC/attention loss \cite{hybridctcattention} for our training process. Let $\mathbf{x} = [x_1,\cdots,x_T]$ be the input frame sequence at the input of Transformer encoder in the fusion module and $\mathbf{y} = [y_1,\cdots, y_L]$ being the targets, where $T$ and $L$ denote the input and target lengths, respectively. 

The CTC loss assumes conditional independence between each output prediction and has a form of

\begin{equation}
p_{\text{CTC}}(\mathbf{y}|\mathbf{x})\approx \prod_{t=1}^{T} p(y_t|\mathbf{x})
\end{equation}

On the other hand, an autoregressive decoder gets rid of this assumption by directly estimating the posterior on the basis of the chain rule, which has a form of 

\begin{equation}
p_{\text{CE}}(\mathbf{y}|\mathbf{x}) = \prod_{l=1}^L p(y_l|y_{<l}, \mathbf{x})
\end{equation}

The overall objective function is computed as follows:

\begin{equation}
\label{eq:lossfunction}
\mathcal{L}=\lambda \log p_{\text{CTC}}(\mathbf{y}|\mathbf{x})+(1-\lambda)\log p_{\text{CE}}(\mathbf{y}|\mathbf{x})
\end{equation}

where $\lambda$ controls the relative weight between CTC loss and seq2seq loss in the hybrid CTC/attention mechanisms. The weight is needed not only when integrating the two losses into one training loss, but also fusing the two predictions during decoding, which we will revisit in the following subsections.


\subsection{Training Pipeline}
The final AVSR model is achieved through a pipeline of training stages.

For audio modality, the audio front-end is first pre-trained through self-supervised learning, which is done by wav2vec 2.0. Then the audio front- and back-end are trained through the audio-only (AO) setting, together with dedicated decoders. 

For the visual modality, the visual front-end is first pre-trained through self-supervised learning, then modified and trained through sequence classification at word level video clips in LRW data. After that, the visual front-end is inherited by the visual-only (VO) model, where visual back-end and dedicated decoders are used. 

The final AVSR model can be trained after the audio-only and visual-only models have converged. Due to computational constraints, we pre-compute the audio and visual back-end outputs, and only learn the parameters in the fusion module and decoders part in this final stage. A detailed visualization of our training pipeline is depicted in Figure \ref{fig:pipeline}.


\subsection{Decoding}
Decoding is performed using joint CTC/attention one-pass decoding \cite{hybridctcattention} with beam search. We apply shallow fusion to incorporate CTC and seq2seq predictions:

\begin{equation}
\begin{aligned}
    \hat{\mathbf{y}}=\mathop{\arg\max}_{\mathbf{y}\in\hat{\mathcal{Y}}}&\{\alpha\log p_{\text{CTC}}(\mathbf{y}|\mathbf{x})\\&+(1-\alpha)\log p_{\text{CE}}(\mathbf{y}|\mathbf{x})\}
\end{aligned}
    \label{eq:lmeq}
\end{equation}
where $\hat{\mathcal{Y}}$ denotes predictions set of target symbols, while $\alpha$ is the relative weight that tuned on validation set. 

\section{Experiments}
In this section, we will first introduce the datasets and various settings we used in each component of our model. Then we will present results of audio-only, visual-only and audio-visual settings. We also present a breakdown of the relative contribution of every component through ablation study. 

\subsection{Dataset}
We use the large-scale publicly AVSR dataset, the Lip Reading Sentences 2 (LRS2) \cite{LRS2} as our main testbed. During training, we also use the Lip Reading in the Wild (LRW) \cite{LRW} as a word-level video classification task to pre-train our visual front-end.

LRS2 consists of 224 hours of aligned audio and videos, with a total of 144K clips from BBC videos, the clips are at a length of sentence level. The training data contains over 2M word instances and a vocabulary of over 40K. The dataset is very challenging as there are large variations in head pose, lighting conditions, genres and the number of speakers.

LRW is a word-level dataset, consisting of 157 hours of aligned audio and videos, totalling 489K video clips from BBC videos, each containing the utterance of a single word out of a vocabulary of 500. The videos have a fixed length of 29 frames, the target word occurring in the middle of the clip and surrounded by co-articulation. All of the videos are either frontal or near-frontal. In our experiment, we only use the visual modality from this dataset to train our visual front-end. 

\subsection{Experimental Settings}
We use character level prediction with an output size of 40, consisting of the 26 characters in the alphabet, the 10 digits, the apostrophe, and special tokens for \texttt{[space]}, \texttt{[blank]} and \texttt{[EOS/SOS]}. Since the transcriptions of the datasets do not contain other punctuations, we do not include them in the vocabulary.

Our implementation is based on the Pytorch library \cite{pytorch} and trained on four NVIDIA A100 GPUs with a total of 160GB memory for 1 week. The network is trained using the Adam optimizer \cite{adam} with $\beta_1=0.9$, $\beta_2=0.999$ and $\epsilon=10^{-8}$ and an initial learning rate of $10^{-4}$. We use label smoothing with a weight set to 0.01, learning rate warm up and reduce on plateau scheduler.  The relative weight in CTC loss and seq2seq loss $\lambda$ is set to 0.2. When decoding, we set $\alpha$ to 0.1. The samples in the pre-train set are cropped by randomly sampling a continuous range of $1/3$ words of the whole utterances, in order to match the length of clips in the train set. Over-length samples are further truncated at 160 frames to reduce memory occupation.
\\[0.5em]
\noindent\textbf{Preprocessing:}
We detected and tracked 68 facial landmarks using dlib \cite{dlib} for each video. To remove differences related to face rotation and scale, the faces are aligned to a neural reference frame using a similarity transformation following \citet{MSTCN}. Interpolation and frame smoothing with a window width of 12 frames are used to deal with the frames that dlib fails to detect. Then a bounding box of $120 \times 120$ is used to crop the mouth ROIs. The cropped frames are further converted to gray-scale and normalized with respect to the overall mean and variance of the train set. Each raw audio waveform is normalized to zero mean and unit variance following \citet{wav2vec2}.
\\[0.5em]
\noindent\textbf{Data Augmentation:}
Following \citet{e2econformer}, random cropping with a size of $112 \times 112$ and horizontal flipping with a probability of 0.5 are performed consistently across all frames of a given image sequence when training visual-only and audio-visual models. For each audio waveform, additive noise is performed in the time domain following \citet{deepavsr} during training audio-only and audio-visual models. Babble noise are added to the audio stream with 5dB SNR and probability of $p_n = 0.25$. The babble noise is synthesized by mixing 20 different audio samples from LRS2.


\begin{table}[t]
    \small
    \centering
\begin{tabularx}{\linewidth}{Xc}
    \toprule
    \makecell{Methods}                             & \makecell{WER}           \\
    \hline
    \hline
    \makecell{Visual-only}                         &                          \\
    \hline
    \makecell{LIBS \cite{lipredingdistill}}        & \makecell{65.3}          \\
    \makecell{TM-CTC* \cite{deepavsr}}             & \makecell{54.7}          \\
    \makecell{Conv-seq2seq \cite{convseq2seq}}     & \makecell{51.7}          \\
    \makecell{TM-seq2seq* \cite{deepavsr}}         & \makecell{50.0}          \\
    \makecell{KD-TM \cite{KDTM}}                   & \makecell{49.2}          \\
    \makecell{LF-MMI TDNN* \cite{overlapped}}      & \makecell{48.9}          \\
    \makecell{E2E Conformer* \cite{e2econformer}}  & \makecell{42.4}          \\
    \makecell{E2E Conformer** \cite{e2econformer}} & \makecell{\textbf{37.9}} \\
    \hline
    \makecell{Our Model}                           & \makecell{43.2}          \\
    \hline
    \hline
    \makecell{Audio-only}                          &                          \\
    \hline
    \makecell{TM-CTC* \cite{deepavsr}}             & \makecell{10.1}          \\
    \makecell{TM-seq2seq* \cite{deepavsr}}         & \makecell{9.7}           \\
    \makecell{CTC/attention* \cite{avsrhybrid}}    & \makecell{8.2}           \\
    \makecell{LF-MMI TDNN* \cite{overlapped}}      & \makecell{6.7}           \\
    \makecell{E2E Conformer** \cite{e2econformer}} & \makecell{3.9}           \\
    \hline
    \makecell{Our Model}                           & \makecell{\textbf{2.7}}  \\
    \hline
    \hline
    \makecell{Audio-Visual}                        &                          \\
    \hline
    \makecell{TM-DCM \cite{TMDCM}}                 & \makecell{8.6}           \\
    \makecell{TM-seq2seq* \cite{deepavsr}}         & \makecell{8.5}           \\
    \makecell{TM-CTC* \cite{deepavsr}}             & \makecell{8.2}           \\
    \makecell{LF-MMI TDNN* \cite{overlapped}}      & \makecell{5.9}           \\
    \makecell{E2E Conformer** \cite{e2econformer}} & \makecell{3.7}           \\
    \hline
    \makecell{Our Model}                           & \makecell{\textbf{2.6}}  \\
    \bottomrule
\end{tabularx}
    \caption{Audio-only, visual-only and audio-visual results of word error rate (WER) tested on LRS2. Models with an * denote that results are using an external language model, which indicates an advantage over our model during evaluation. Models denoted with ** means that it uses a more powerful Transformer language model.
    }
    \label{tb:lrs2result}
\end{table}

\begin{table}[ht]
    \small
    \centering
\begin{tabularx}{\linewidth}{ccXX}
    \toprule
    Modules          & \makecell{Ours}    & \makecell{TM-CTC}    & \makecell{E2E\\Conformer}       \\
    \hline
    Audio front-end  & \makecell{315.0M} & \makecell{-} & \makecell{3.9M} \\
    \hline
    Visual front-end & \makecell{23.5M} & \makecell{11.2M\\(freezed)} & \makecell{11.2M} \\
    \hline
    Audio back-end & \makecell{20.2M} & \makecell{20.2M} & \makecell{31.8M} \\
    \hline
    Visual back-end & \makecell{20.2M} & \makecell{20.2M} & \makecell{31.8M} \\
    \hline
    Fusion module & \makecell{19.7M} & \makecell{19.7M} & \makecell{0.8M} \\
    \hline
    Decoders      & \makecell{26.2M} & \makecell{20.5K} & \makecell{9.5M} \\
    \bottomrule
\end{tabularx}
    \caption{The parameters comparison of ours, TM-CTC \cite{deepavsr} and E2E Conformer \cite{e2econformer} models.}
    \label{tb:modelsize}
\end{table}

\noindent\textbf{Evaluation:}
For all experiments, word error rate (WER) are reported which is defined as $\text{WER} = (S + D + I)/N$. The $S$, $D$ and $I$ in the formula denotes the number of substitutions, deletions and insertions respectively from the reference to the hypothesis, and $N$ is the number of words in the inference. The babble noise added to the audio waveform during evaluation is generated using the same manner as training, while we set a different seed to avoid model fit to a specific generated noise. Decoding is performed using joint CTC/attention one-pass decoding \cite{hybridctcattention} with beam width 5 (the values were determined on the held-out validation set of LRS2). We don't use an external language model in our experiments.

\subsection{Results}
We present results for all experiments in Table \ref{tb:lrs2result}, reporting WERs on visual-only, audio-only and audio-visual models. Note that many of the models listed here are also using extra training data in different stages of training pipeline, such as MV-LRS \cite{chung2017lip}, LRS3 \cite{LRS3}, LibriSpeech \cite{librispeech} and LRW. 

We present the parameters of our model, TM-CTC model \cite{deepavsr} and the current state-of-the-art model \cite{e2econformer} in Table \ref{tb:modelsize}. Our model back-ends and fusion module configurations follow TM-CTC model, the hyper-parameters settings in the seq2seq decoder are the same as in the back-ends. The most significant difference is that we utilize pre-trained front-ends, resulting in a larger model size.
\\[0.5em]
\noindent\textbf{Audio-visual Setting:}
In the main audio-visual setting, the pre-train and train sets in LRS2 are used as train set in the final training stage. 
Our proposed audio-visual model achieves a WER of 2.6\% without the help of an external language model, which improves by 1.1\% over the current state-of-the-art \cite{e2econformer}. This is rather a big improvement, with a relative improvement of around 30\%.
\\[0.5em]
\noindent\textbf{Audio-only Setting:}
The training data used for training audio-only model consists of 224 hours labelled data from LRS2, as well as the 60K hours unlabelled data from LibriLight \cite{librilight} that are indirectly used through inheriting wav2vec 2.0 parameters.
Our model also achieves a WER of 2.7\%, which reduces the WER of the current state-of-the-art \cite{e2econformer} by 1.2\%, indicating a relative improvement of 31\%. 
\\[0.5em]
\noindent\textbf{Visual-only Setting:}
The visual-only model uses labelled LRS2 data in its pre-train and train sets, the LRW for supervised pre-training, and indirectly using the 1.28M unlabelled images from ImageNet through MoCo v2. The visual-only model achieves a WER of 43.2\%, lagging behind the current state-of-the-art E2E Conformer model \cite{e2econformer} with 5.3\%. 
Compared to E2E Conformer, the main difference is that a large Transformer language model is used during decoding, which itself brings a 4.5\% difference compared with a normal RNN language model in their ablation studies \cite{e2econformer}. The gap between our visual-only model and the E2E Conformer model with a RNN language model is 0.8\%, which resides in a quite reasonable range. Additionally, we use a 6-layers Transformer encoder for temporal modelling instead of a 12-layers conformer encoder, which resulted in a smaller back-end size. 

If we consider a fairer comparison by only looking at benchmarks without using an external language model, the best-reported benchmark is \citet{KDTM}, which achieved a WER of 49.2\%, lagging behind our model by 6.0\%.


\subsection{Ablation Studies}
In this section, we investigate the impact of every individual building block by testing them in LRW, audio-only and visual-only settings.
\\[0.5em]
\noindent\textbf{MoCo v2 Contribution in Visual Word Classification:}
Results of visual word classification on LRW are shown in Table \ref{tb:lrwablation}. We first train a model by replacing the ResNet-18 front-end in \citet{LSTMlipreading} with a ResNet-50 front-end, matching the size of MoCo v2 but with fresh weights. This results in an absolute improvement of 2.1\%. Then we initialize the ResNet-50 front-end with MoCo v2 weights and a further absolute improvement of 2.3\% is observed, which implies that self-supervised learning is actually functioning in better represent the lip movement. Additionally, When Using 6 layers of Transformer encoder instead of TCN as back-end, we can observe another absolute improvement of 6.0\%. We also noticed that using MoCo v2 front-end could significantly reduce the training time. 

\begin{table}[!h]
    \small
    \centering
    \def\arraystretch{1}
    \begin{tabularx}{\linewidth}{Xc}
        \toprule
        Method & Acc\\
        
        \midrule
        Baseline\cite{LSTMlipreading} & 74.6\%\\
        \qquad+ ResNet-50 front-end & 76.7\%\\
        \qquad\qquad+ MoCo v2 front-end & 79.0\%\\
        \qquad\qquad\qquad+ Transformer encoder back-end & \textbf{85.0}\%\\
        \bottomrule
    \end{tabularx}
    \caption{Ablation study on visual word classification performance on LRW.}
    \label{tb:lrwablation}
\end{table}
\noindent\textbf{Performance Breakdown in Audio-only Setting:}
Results of audio-only model on LRS2 are shown in Table \ref{tb:audioablation}. Starting from \citet{deepavsr}, we first train a model by replacing the STFT audio feature with a wav2vec 2.0 front-end pre-trained on LibriSpeech, resulting in an absolute improvement of 11.1\%. Then we use another pre-trained model learned on an even larger unlabelled single modality dataset Libri-Light, and a further absolute improvement of 0.6\% is observed. We further train the model with a hybrid CTC/attention decoder during the training stage, which results in another absolute improvement of 0.9\%.

\begin{table}[!h]
    \small
    \centering
    \def\arraystretch{1}
    \begin{tabularx}{\linewidth}{Xc}
        \toprule
        Method & WER\\
        
        \midrule
        Baseline\cite{deepavsr} & 15.3\%\\
        \qquad+ wav2vec 2.0 (LibriSpeech) encoder & 4.2\%\\
        \qquad\qquad+ wav2vec 2.0 (LibriLight) encoder & 3.6\%\\
        \qquad\qquad\qquad+ Hybrid CTC/attention & \textbf{2.7}\%\\
        \bottomrule
    \end{tabularx}
    \caption{Ablation study on audio-only model performance on LRS2.}
    \label{tb:audioablation}
\end{table}
\noindent\textbf{Performance Breakdown in Visual-only Setting:}
Results of the visual-only model on LRS2 are shown in Table \ref{tb:visualablation}. Starting from \citet{deepavsr}, we first introduce end-to-end training by using a hybrid CTC/attention decoder (the front-end is still pre-trained through LRW), resulting in an absolute improvement of 16.0\%. Then we initialize the front-end with pre-trained MoCo v2 weights, a same end-to-end training manner results in a further absolute improvement of 5.8\%.

\begin{table}[!h]
    \small
    \centering
    \def\arraystretch{1}
    \begin{tabularx}{\linewidth}{Xc}
        \toprule
        Method & WER\\
        
        \midrule
        Baseline\cite{deepavsr} & 65.0\%\\
        \qquad+ Hybrid CTC/attention & 49.0\%\\
        \qquad\qquad+ MoCo v2 front-end & \textbf{43.2}\%\\
        \bottomrule
    \end{tabularx}
    \caption{Ablation study on visual-only model performance on LRS2.}
    \label{tb:visualablation}
\end{table}

\noindent\textbf{Robustness under Noisy Inputs:}
To evaluate the model's tolerance to audio noise, we tested the performance of our model under babble noise with different SNR levels. Our audio-only and audio-visual models reach WERs of 32.5\% and 24.5\% when the SNR level is 0dB, respectively, which reduce the reported result in \citet{deepavsr} by 25.5\% and 9\%\footnote{\citet{e2econformer} also provides a performance under noisy inputs, however, we are not able to compare with them due to a lack of necessary details to generate the same noise. }. When the SNR level rises to 5dB, our audio-only and audio-visual model obtain WERs of 6.8\% and 6.3\%.

\begin{figure}[t]
    \centering
    \includegraphics[width=\columnwidth]{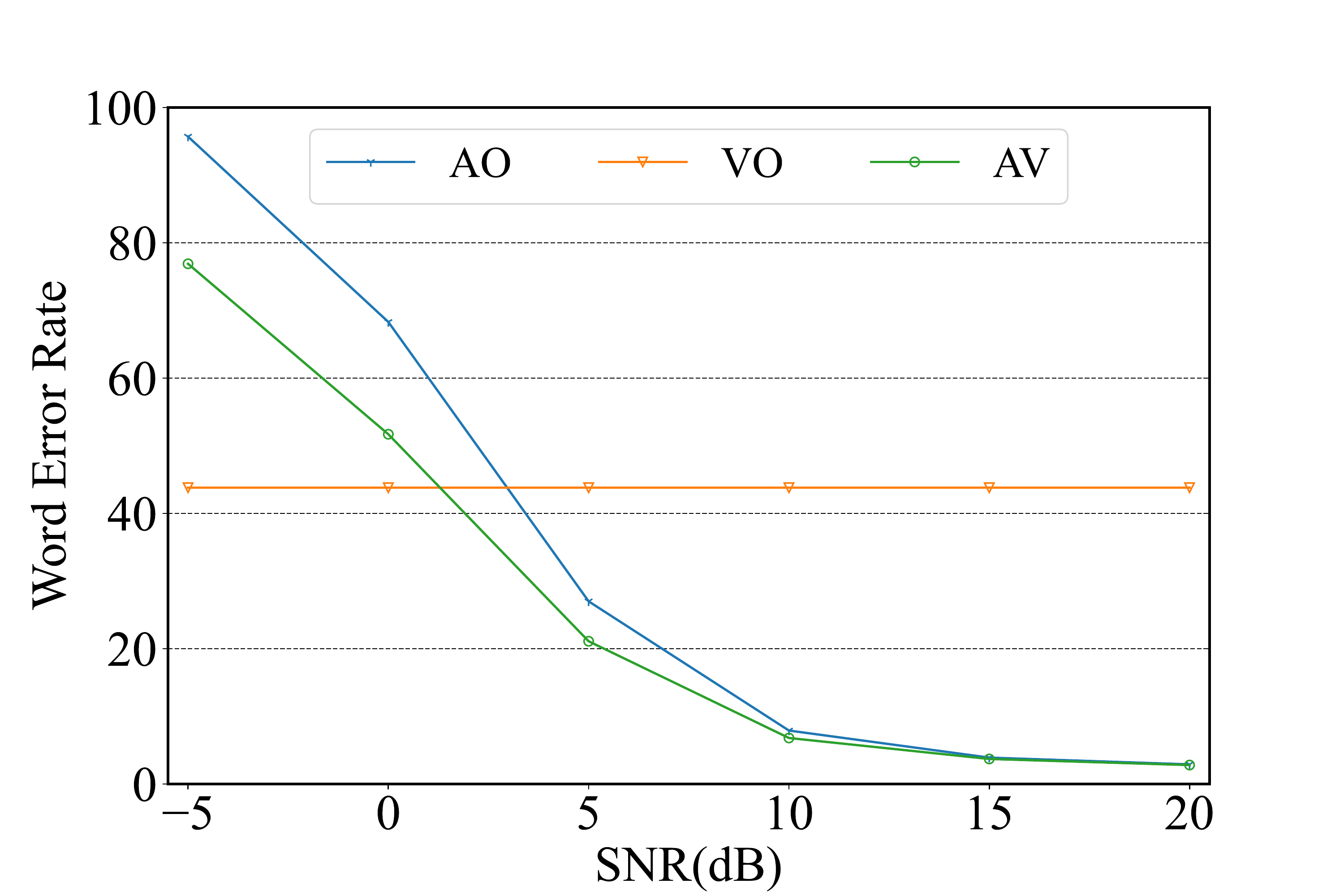}
    \caption{Word error rate (WER) under different SNR levels. The noises are human speech sampled from LRS2. AO: Audio-Only model, VO: Visual-Only model, AV:Audio-Visual model}
    \label{fig:human_noise}
\end{figure}

Besides achieving significant improvement over the baseline model under babble noise environment, we further investigate the model performance under human noise environment. The human noise is extremely challenging because the noise itself contains some words, while the model cannot easily distinguish which audio signal is the one to be recognized. We synthesize the human noise by randomly crop many 1 second signals from different audio samples in the LRS2 dataset. As shown in Fig. \ref{fig:human_noise}, we conduct experiments varying different levels of human noise, the models are trained using babble noise augmented audio. The WER increases greatly after the SNR level drops down under 0db. It is because the model may not be able to distinguish the two overlapped spoken words at a low SNR level.

And the overall performance under each SNR level is worse than babble noise, indicating that noise with specific information is harder than disorganized babble noise.

\begin{table}[t]
    \small
    \centering
\begin{tabularx}{\linewidth}{c|c|XXX}
    \toprule
    Modality          & Model            & 0dB             & 5dB   & clean          \\
    \hline
    \multirow{2}*{AO} & \citet{deepavsr} & 58.0\%          & -     & 10.5\%         \\
    \cline{2-5}
    ~                 & Our model        & \textbf{32.5}\% & 6.8\% & \textbf{2.7}\% \\
    \hline
    \multirow{2}*{AV} & \citet{deepavsr} & 33.5\%          & -     & 9.4\%          \\
    \cline{2-5}
    ~                 & Our model        & \textbf{24.5}\% & 6.3\% & \textbf{2.6}\% \\
    \bottomrule
\end{tabularx}
    \caption{Word error rate (WER) under different SNR levels. The noises are synthesized babble noises.}
    \label{tb:noiseperformance}
\end{table}

\noindent\textbf{Recognition under Low Resource:}
A significant benefit of using self-supervised pre-trained models is that only a small amount of labelled data is needed for training a model. To further investigate the models' performance in low resource environment, we use the 28 hours train set of LRS2 to train an audio-only and a visual-only model. The results are shown in Table \ref{tb:lowresource}. The audio-only model trained with 28 hours data achieves a WER of 3.4\%, which is a little bit worse than the one trained with 224 hours data. The result indicates that for the audio-only model, the self-supervised model pre-trained on a large-scale single modality dataset can significantly reduce the demands of data. While the visual-only model trained with 28 hours data has a great gap with the one trained with 224 hours data, the reason can be that the visual-only model is harder to train and demands a larger amount of data.

\begin{table}[t]
    \small
    \centering
\begin{tabularx}{\linewidth}{c|c|X}
    \toprule
    \makecell{Model}                      & \makecell{Training data (Hours)} & \makecell{WER (\%)}        \\
    \hline
    \makecell{\multirow{2}*{audio-only}}  & \makecell{LRS2 (224)}            & \makecell{2.7}        \\
    \cline{2-3}
    ~                                     & \makecell{LRS2 train set (28)}    & \makecell{3.4 (+0.7)}\\
    \hline
    \makecell{\multirow{2}*{visual-only}} & \makecell{LRS2 (224)}            & \makecell{43.2}       \\
    \cline{2-3}
    ~                                     & \makecell{LRS2 train set (28)}   & \makecell{68.9 (+25.7)} \\
    \bottomrule
\end{tabularx}
    \caption{Performance of audio-only and visual-only models using different training data.}
    \label{tb:lowresource}
\end{table}


\subsection{Discussion and Conclusion}
In this work, we propose to utilize self-supervised learning for AVSR by simply incorporating the pre-trained model trained in massive unlabelled single modality data. Although the visual pre-trained models are not straight-forward to be transplanted into visual front-end, we still manage to integrate pre-trained models in both modalities for the AVSR task. Experimental results are impressive, resulting in a 30\% relative improvement.

It's interesting to observe that self-supervised model in audio modality has an even larger improvement than that of the visual counterpart. We believe the reasons can be listed as follows:
\begin{itemize}
    \item The training data scale of audio modality is significantly larger than that of visual modality, with the Libri-Light dataset used for pre-training wav2vec 2.0 consists of 60K hours audio signals, the ImageNet dataset, on the contrary, has only 1.28M images, roughly equivalent to 14 hours silent video under 25 FPS.
    \item The MoCo v2 model is pre-trained on images to better represent frame-level contents, while there are no pre-training steps to model the temporal correlation between frames. In contrast, the wav2vec 2.0 model is pre-trained on consistent audios, thus having a better temporal modelling ability.
\end{itemize}



As there has not emerged a dominating cross-modality self-supervised learning approach in the field of AVSR, in future work, we are going to explore two more directions in the self-supervised learning scenario based on this work. The first is utilizing the temporal correlations within the visual domain, while the other is the cross-modal correlations between the audio and visual modality. We hope this work could pave the way towards multimodality self-supervised learning, especially for various aspects in 
AVSR. 

\section*{Ethical Statement}
This work will not pose ethical problems, the data resources we use are all from published works and do not involve privacy issues related to data collection. The data is collected from BBC and contains thousands of diverse speakers, allowing the speech recognition models to generalize to all speakers. In terms of computational experiments, we used publicly available pre-trained models, which makes the training more environmentally friendly and lowers the computational requirements to reproduce our work.

\section*{Acknowledgements}
This work was sponsored by the National Natural Science Foundation of China (NSFC) grant (No. 62106143), and Shanghai Pujiang Program. We would like to thank all the anonymous reviewers for their valuable and constructive comments.

\bibliographystyle{acl_natbib}
\bibliography{anthology,custom}

\clearpage
\appendix

\section{Decoding Algorithm}

\begin{algorithm}[ht!]
\small
\caption{Hybrid CTC/attention one-pass decoding adapted from \citet{hybridctcattention}. Notation: $X$ is the speech input; $L_{max}$ is the maximum length of the hypotheses to be searched, we set it to $T$; $C$ is the decoded symbol sequence; $\texttt{[b]}$ denotes $\texttt{[blank]}$.}
\label{alg:decode}
    \textbf{Input}: $X, L_{max}$\\
    \textbf{Output}: $C$
        \begin{algorithmic}[1] 
        \STATE $\Omega_0=\texttt{\{[SOS]\}}$
        \STATE $\hat{\Omega}=\emptyset$
        \STATE $\gamma_0^{(b)}(\texttt{[SOS]})=1$
        \FOR{$t=1,\cdots,T$}
            \STATE $\gamma_t^{(n)}(\texttt{[SOS]})=0$
            \STATE $\gamma_t^{(b)}(\texttt{[SOS]})=\prod\limits^t_{\tau=1}\gamma_{\tau-1}^{(b)}(\texttt{[SOS]})\cdot p(z_\tau=\texttt{[b]}|X)$
        \ENDFOR
        \FOR{$l=1\cdots L_{max}$}
            \STATE $\Omega_l=\emptyset$
            \WHILE{$\Omega_{l-1}\neq\emptyset$}
                \STATE $g=\text{HEAD}(\Omega_{l-1})$
                \STATE $\text{DEQUEUE}(\Omega_{l-1})$
                \FOR{each $c \in \mathcal{U}$}
                    \STATE $h=g\cdot c$
                    \IF {$c=\texttt{[EOS]}$}
                        \STATE $\log p_{\text{ctc}}(h|X)=\log \{\gamma_T^{(n)}(g)+\gamma_T^{(b)}(g)\}$
                    \ELSE
                        \IF {$g=\texttt{[SOS]}$}
                            \STATE $\gamma_1^{(n)}(h)=p(z_1=c|X)$
                        \ELSE
                            \STATE $\gamma_1^{(n)}(h)=0$
                        \ENDIF
                        \STATE $\gamma_1^{(b)}(h)=0$
                        \STATE $\Psi = \gamma_1^{(n)}(h)$
                        \FOR{$t=2\cdots T$}
                            \IF {$last(g)=c$}
                                \STATE $\Phi=\gamma_{t-1}^{(b)}(g)$
                            \ELSE
                                \STATE $\Phi=\gamma_{t-1}^{(b)}(g) + \gamma_{t-1}^{(n)}(g)$
                            \ENDIF
                            \STATE $\gamma_{t}^{(n)}(h) = (\gamma_{t-1}^{(n)}(h) + \Phi)p(z_t=c|X)$
                            \STATE $\gamma_{t}^{(b)}(h) = (\gamma_{t-1}^{(b)}(h) + \gamma_{t-1}^{(n)}(h))p(z_t=\texttt{[b]}|X)$
                            \STATE $\Psi = \Psi + \Phi\cdot p(z_t=c|X)$
                        \ENDFOR
                        \STATE $\log p_{\text{ctc}}(h|X)=\log (\Psi)$
                    \ENDIF
                    \STATE $\log p(h|X) = \alpha\log p_{\text{ctc}}(h|X)$\newline
                    \hspace*{5em}$+ (1-\alpha)\log p_{\text{att}}(h|X)$
                    \IF {$c=\texttt{[EOS]}$}
                        \STATE $\text{ENQUEUE}(\hat{\Omega},h)$
                    \ELSE
                        \STATE $\text{ENQUEUE}(\Omega_l,h)$
                    \ENDIF
                \ENDFOR
            \ENDWHILE
            \STATE $\Omega_l = \text{TOPK}(\Omega_l, W)$
        \ENDFOR
        \STATE \textbf{return} $\text{arg max}_{C\in\hat{\Omega}} \log p(C|X)$
        \end{algorithmic}
\end{algorithm}

Algorithm \ref{alg:decode} describes the hybrid CTC/attention decoding procedure. The CTC prefix probability is defined as the cumulative probability of all label sequences that have $h$ as their prefix:
\begin{equation}
    p_{\text{ctc}}(h|X)=\sum\limits_{v\in (\mathcal{U})^+}p_{\text{ctc}}(h\cdot v|X)
\end{equation}
where $v$ denotes all possible symbol sequences except the empty. The CTC probability can be computed by keeping the forward hypothesis probabilities $\gamma_t^{(n)}$ and $\gamma_t^{(b)}$, where the superscripts $(n)$ and $(b)$ represents all CTC paths end with a non-$\texttt{[blank]}$ or $\texttt{[blank]}$ symbol, respectively.

The decoding algorithm is also a beam search with width $W$ and hyperparameter $\alpha$ control the relative weight given to CTC and attention decoding. $\mathcal{U}$ is a set of symbols excluding $\texttt{[blank]}$, and a same token is used to represent $\texttt{[SOS]}$ and $\texttt{[EOS]}$ in our implementation.

\section{Decoding Examples}
\begin{table}[h!]
\small
\begin{tabularx}{\linewidth}{X}
    \toprule
    \textit{\texttt{AO:}} WHATEVER YOU \underline{ASK}\\
    \textit{\texttt{AV:}} WHATEVER YOU ARE\\
    \hline
    \textit{\texttt{AO:}} TRAVEL THREE MILES \underline{URBER} WEST AND YOU DO GET MORE FOR YOUR MONEY HERE\\
    \textit{\texttt{AV:}} TRAVEL THREE MILES FURTHER WEST AND YOU DO GET MORE FOR YOUR MONEY HERE\\
    \hline
    \textit{\texttt{AO:}} IT COULD BE YOUR PASSPORT \underline{FOR} A SMALL FORTUNE\\
    \textit{\texttt{AV:}} IT COULD BE YOUR PASSPORT TO A SMALL FORTUNE\\
    \hline
    \textit{\texttt{AO:}} \underline{WHAT} TO THINK FOR THEMSELVES\\
    \textit{\texttt{AV:}} NOT TO THINK FOR THEMSELVES\\
    \hline
    \textit{\texttt{AO:}} NOT \underline{THE} SUBJECT \underline{MATTERING}\\
    \textit{\texttt{AV:}} NOT FOR SUBJECT MATTER\\
    \hline
    \textit{\texttt{AO:}} I WOULDN'T SAY I'M \underline{THE} STAR\\
    \textit{\texttt{AV:}} I WOULDN'T SAY I'M A STAR\\
    \hline
    \textit{\texttt{AO:}} \underline{CRISPAS} PUDDING THAT NOBODY REALLY LIKES\\
    \textit{\texttt{AV:}} CHRISTMAS PUDDING THAT NOBODY REALLY LIKES\\
    \hline
    \textit{\texttt{AO:}} \underline{BUT} AT THE SAME TIME\\
    \textit{\texttt{AV:}} AT THE SAME TIME\\
    \hline
    \textit{\texttt{AO:}} BEING \sout{ON} MY OWN\\
    \textit{\texttt{AV:}} BEING MY OWN\\
    \hline
    \textit{\texttt{AO:}} \sout{SO} AT ONE POINT\\
    \textit{\texttt{AV:}} AT ONE POINT\\
    \bottomrule
\end{tabularx}
    \caption{AO (audio-only) and AV (audio-visual) decoding examples. Underline denotes substitutions and insertions error; Strikethrough denotes deletions error.}
    \label{tb:AOAV_match}
\end{table}

Table \ref{tb:AOAV_match} is examples of sentences that audio-only model fails to predict while audio-visual model correctly predicts. The visual modality enhances the model from a wide range of error cases. 

\section{Preprocessing Example}

\begin{figure*}[!ht]
    \centering
    \subfloat[Landmarks detected by dlib. Green dots are 68 landmarks, frames without landmarks are ones that dlib fail to detect.]
    {
        \includegraphics[width=\textwidth]{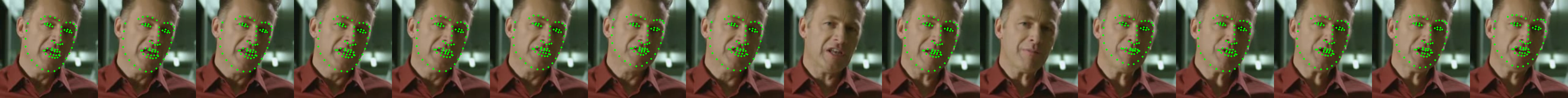}
        \label{fig:detect}
    }\\
    \subfloat[Landmarks after linear interpolation.]
    {
        \includegraphics[width=\textwidth]{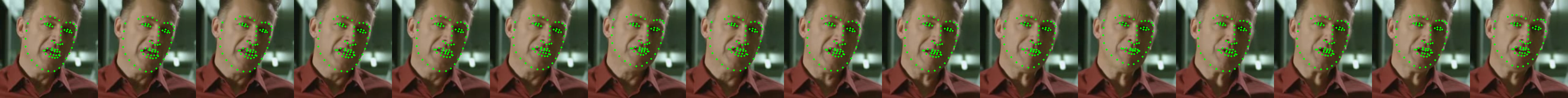}
        \label{fig:interpolate}
    }\\
    \subfloat[Faces smoothed with a window width of 12 and aligned to a neural reference frame using a similarity transformation.]
    {
        \includegraphics[width=\textwidth]{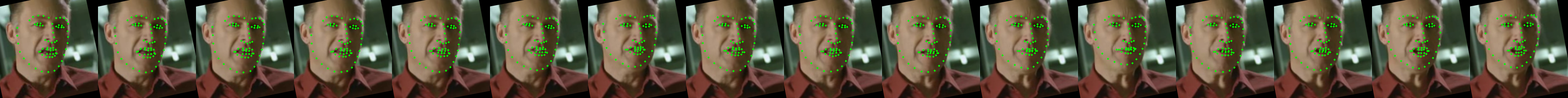}
        \label{fig:transpose}
    }\\
    \subfloat[Mouth ROIs cropped using a bounding box of $120\times 120$.]
    {
        \includegraphics[width=\textwidth]{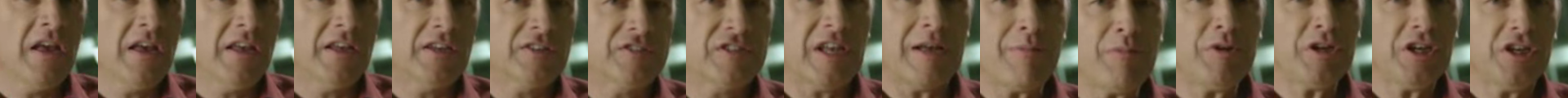}
        \label{fig:cut}
    }
    \caption{Preprocessing example to illustrate the process to generate mouth ROIs.}
    \label{fig:preprocessexample}
\end{figure*}

The input images are sampled at 25 FPS and resized to $224\times 224$ pixels. We crop a $120\times 120$ mouth ROI from each frame. Fig. \ref{fig:preprocessexample} shows the process to generate. 

\end{document}